\documentclass[10pt,twocolumn,letterpaper]{article}

\usepackage{cvpr}
\usepackage{times}
\usepackage{epsfig}
\usepackage{graphicx}
\usepackage{amsmath}
\usepackage{amssymb}
\makeatletter
\@namedef{ver@everyshi.sty}{}
\makeatother
\usepackage{tikz}

\usepackage{amsmath,amsfonts,bm}

\def\eqref#1{equation~\ref{#1}}

\def\1{\bm{1}}

\def\va{{\bm{a}}}

\def\vc{{\bm{c}}}

\def\vf{{\bm{f}}}

\def\vl{{\bm{l}}}

\def\vn{{\bm{n}}}

\def\vs{{\bm{s}}}
\def\vt{{\bm{t}}}

\def\vv{{\bm{v}}}

\DeclareMathAlphabet{\mathsfit}{\encodingdefault}{\sfdefault}{m}{sl}
\SetMathAlphabet{\mathsfit}{bold}{\encodingdefault}{\sfdefault}{bx}{n}

\def\gL{{\mathcal{L}}}

\def\sN{{\mathbb{N}}}

\def\sR{{\mathbb{R}}}

\usepackage{url}
\usepackage{makecell} 
\usepackage{pdfpages}
\usepackage{adjustbox}

\usepackage[utf8]{inputenc}

\usepackage[numbers,sort,compress]{natbib}
\usepackage{multirow}
\usepackage[utf8]{inputenc} %
\usepackage[T1]{fontenc}    %
\usepackage{booktabs} %
\usepackage{balance} %
\usepackage{amssymb}
\usepackage{subcaption}
\usepackage{array} 
\usepackage{multirow}
\usepackage[linesnumbered,ruled]{algorithm2e}
\usepackage{algpseudocode}
\usepackage{amsmath,amssymb,mathtools}
\usepackage{cuted}
\usepackage{amsthm,bm}
\usepackage{afterpage}
\usepackage{bbm}
\usepackage{color}
\usepackage{algorithm2e}
\usepackage{mathtools}

\newcommand{\adv}{\mathrm{adv}}
\newcommand{\perc}{\mathrm{perceptual}}
\usepackage{xspace}

\SetCommentSty{mycommfont}

\newcommand{\reffig}[1]{Figure~\ref{fig:#1}}
\newcommand{\refsec}[1]{Section~\ref{sec:#1}}

\newcommand{\reftbl}[1]{Table~\ref{tbl:#1}}

\newcommand{\refeqlong}[1]{Equation~\ref{eq:#1}}

\newcommand{\refeqshort}[1]{(\ref{eq:#1})}

\newcommand{\lblfig}[1]{\label{fig:#1}}

\newcommand{\ignorethis}[1]{}
\newcommand{\norm}[1]{\lVert#1\rVert}

\usepackage{subcaption}
\usepackage{array}
\usepackage{multirow}
\usepackage[linesnumbered,ruled]{algorithm2e}
\usepackage{algpseudocode}
\usepackage{amsmath,amssymb,mathtools}
\usepackage{cuted}
\usepackage{amsthm}
\usepackage{afterpage}
\usepackage{xcolor}

\newif\ifsubmit

\submitfalse

\ifsubmit
\newcommand{\bo}[1]{\textcolor{blue}{}}
\newcommand{\chaowei}[1]{\textcolor{cyan}{}}
\newcommand{\dawei}[1]{\textcolor{yellow}{}}
\newcommand{\mingyan}[1]{\textcolor{purple}{}}
\newcommand{\jia}[1]{\textcolor{green}{}}
\else
\newcommand{\bo}[1]{\textcolor{blue}{Bo: #1}}
\newcommand{\chaowei}[1]{\textcolor{cyan}{Chaowei: #1}}
\newcommand{\dawei}[1]{\textcolor{orange}{Dawei: #1}}
\newcommand{\mingyan}[1]{\textcolor{purple}{Mingyan: #1}}
\newcommand{\jia}[1]{\textcolor{green}{Jia: #1}}
\fi
\newcommand{\StAdv}{\textit{MeshAdv}\xspace}
\newcommand{\stAdv}{\textit{meshAdv}\xspace}
\newcommand{\densenet}{DenseNet\xspace}
\newcommand{\inception}{Inception-v3\xspace}
\newcommand{\yolo}{Yolo-v3\xspace}

\newcommand{\pascal}{PASCAL3D+\xspace}
\newcommand{\unknown}{{*}}
\makeatletter
\renewcommand{\paragraph}{%
  \@startsection{paragraph}{4}%
  {\z@}{1.2ex \@plus 1ex \@minus .2ex}{-1em}%
  {\normalfont\normalsize\bfseries}%
}
\makeatother
\usepackage[pagebackref=true,breaklinks=true,letterpaper=true,colorlinks,bookmarks=false]{hyperref}

\cvprfinalcopy %

\def\cvprPaperID{2440} %

\ifcvprfinal\fi
\begin{document}

\title{MeshAdv: Adversarial Meshes for Visual Recognition}
\author{Chaowei Xiao \thanks{Alphabetical ordering; The first two authors contributed equally.}$^{\ \,1}$ \quad Dawei Yang $^{*1,2}$  \quad  Jia Deng $^{2}$ \quad Mingyan Liu $^1$\quad  Bo Li $^{3}$\\
$^1$ University of Michigan, Ann Arbor\\
$^2$ Princeton University\\
$^3$ UIUC}

\maketitle
\thispagestyle{empty}
\begin{abstract}
Highly expressive models such as deep neural networks (DNNs) have been widely applied to various applications. 
However, recent studies show that DNNs are vulnerable to adversarial examples, which are carefully crafted inputs aiming to mislead the predictions.
Currently, the majority of these studies have focused on perturbation added to image pixels, while such manipulation is not physically realistic. Some works have tried to overcome this limitation by attaching printable 2D patches or painting patterns onto surfaces, but can be potentially defended because 3D shape features are intact. %
In this paper, we propose \stAdv to generate ``adversarial 3D meshes'' from objects that have rich shape features but minimal textural variation. %
To manipulate the shape or texture of the objects, we make use of a differentiable renderer to compute accurate shading on the shape and propagate the gradient.
Extensive experiments show that the generated 3D meshes are effective in attacking both classifiers and object detectors.
We evaluate the attack under different viewpoints.
In addition, we design a pipeline to perform black-box attack on a photorealistic renderer with unknown rendering parameters.

\end{abstract}

\section{Introduction}
Despite the increasing successes in various domains~\citep{he2016deep,collobert2008unified,deng2013recent,silver2016mastering}, deep neural networks (DNNs) are found vulnerable to adversarial examples:
a deliberate perturbation of small magnitude on the input can make a network output incorrect predictions.
Such adversarial examples have been widely studied in 2D domain~\citep{szegedy2013intriguing,goodfellow2014explaining,moosavi2016deepfool,papernot2016limitations,carlini2017towards,xiao2018generating,xiao2018spatially,xiao2018characterizing}, but the attack generated by directly manipulating pixels can be defended by securing the camera, so that the generated images are not realizable in practice.
To overcome this issue, there has been significant prior progress on generating physically possible adversarial examples~\citep{kurakin2016adversarial,evtimov2017robust,athalye2017synthesizing,brown2017adversarial} by altering the texture of a 3D surface, \ie applying adversarial printable 2D patches or painting patterns. %
Such attacks, however, are less suitable for textureless objects, because adding texture to an otherwise textureless surface may increase the chance of being detected and defended.  

\begin{figure}[t]
    \vspace{27pt}
    \centering
    \includegraphics[width=\linewidth]{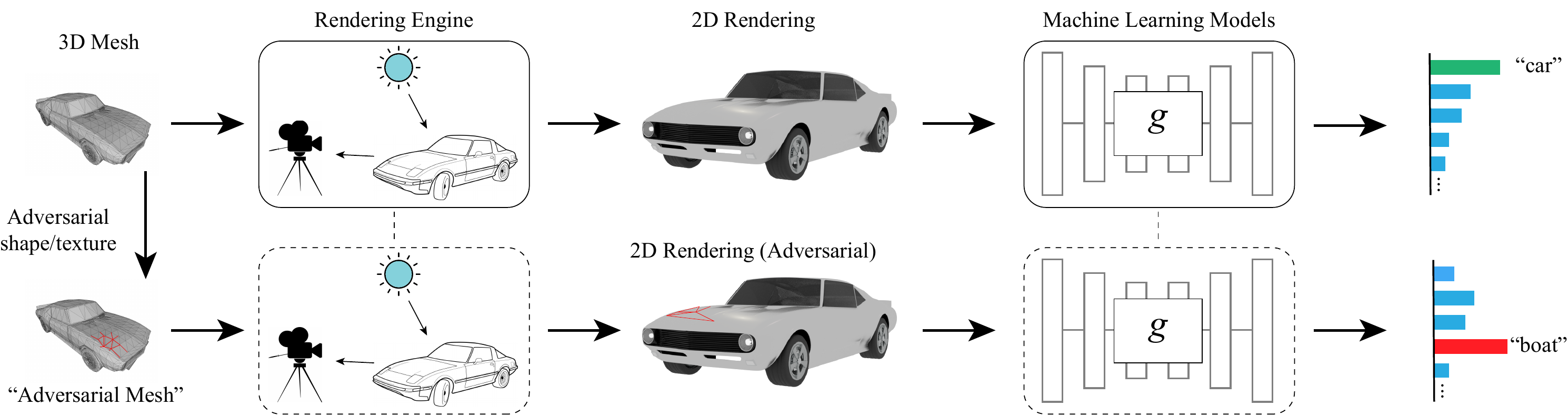}
    \caption{The pipeline of ``adversarial mesh'' generation by \stAdv.}
    \label{fig:pull_fig}
\end{figure}

In this work, we explore a new avenue of attack where we generate physically possible adversarial examples by altering 3D shape. We explore 3D objects that have rich shape features but minimal texture variation, and show that we can still fulfill the adversarial goal by perturbing the shape of those 3D objects, while the same method can still be applied to textures.
Specifically,
we propose \stAdv to generate adversarial meshes with negligible perturbations.
We leverage a physically based differentiable renderer~\citep{kato2018renderer} %
to accurately render the mesh under certain camera and lighting parameters. A deep network then outputs a prediction given the rendered image as input.
Since this whole process is differentiable, gradients can be propagated from the network prediction back to the shape or texture of the mesh. Therefore, we can use gradient based optimization to generate shape based or texture based perturbation by applying losses on the network output. The entire pipeline is shown in~\reffig{pull_fig}.

Even though we are only manipulating physical properties (shape and texture) of a 3D object, we can always fool state of the art DNNs (see \refsec{pascal_classification}). Specifically, we show that for any fixed rendering conditions (\ie lighting and camera parameters), state of the art object classifiers (\densenet~\citep{huang2017densely} and \inception~\citep{szegedy2016rethinking}) and detector (\yolo~\citep{redmon2018yolov3}) can be consistently tricked by slightly perturbing the shape and texture of 3D objects.
We further show that by using multiple views optimization, the attack success rate of ``adversarial meshes'' increases under various viewpoints (see \reftbl{multiview}).
In addition, we conduct user studies to show that the generated perturbation are negligible to human perception.

Since the perturbation on meshes is adversarially optimized with the help of a differentiable renderer, a natural question to ask is whether a similar method can be applied in practice when the rendering operation is not differentiable. %
We try to answer this question by proposing a pipeline to perform black-box attack on a photorealistic renderer (with non-differentiable rendering operation) under unknown rendering parameters.
We show that via estimating the rendering parameters and improving the robustness of perturbation, our generated ``adversarial meshes'' can attack on a photorealistic renderer. %

Additionally, we visualize our shape based perturbation to show possible vulnerable regions for meshes. This can be beneficial when we hope to improve the robustness (against shape deformation) of machine learning models that are trained on 3D meshes~\citep{change2015shapenet,wu3dshapenets,song2016ssc} for different tasks such as view point
estimation~\citep{su2015render4cnn}, indoor scene 
understanding~\citep{song2016ssc,zhang2016physically,handa2016understanding,mccormac2017scenenet} and so on~\citep{richter2016playing,yang2018shape,massa2016deep,varol2017surreal,chen2015deep3dpose}.

To summarize, our {\em contributions} are listed below: 
1) We propose an end-to-end optimization based method \stAdv to generate 3D ``adversarial meshes'' with negligible perturbations, and show that it is effective in attacking different machine learning tasks;
2) We demonstrate the effectiveness of our method on a black-box non-differentiable renderer with unknown parameters;
3) We provide insights into vulnerable regions of a mesh via visualizing the flow of shape based perturbations;
4) We conduct user studies to show that our 3D perturbation is subtle enough and will not affect user recognition.

\section{Related Work}
\paragraph{Adversarial Attacks}
Adversarial examples have been heavily explored in 2D domains~\citep{szegedy2013intriguing,goodfellow2014explaining,moosavi2016deepfool,papernot2016limitations,xiao2018generating,xiao2018spatially}, but directly manipulation of image pixels requires access to cameras. %
To avoid this, physical adversarial examples studied in~\citep{kurakin2016adversarial,evtimov2017robust} show impressive robust adversarial examples under camera transformations. However, the perturbations are textured based and may not be applied to arbitrary 3D shapes. %

Meanwhile, \citet{athalye2017synthesizing} further advance texture based adversarial examples by enhancing the robustness against color transformations, and show that the generated textures for a turtle and a baseball that can make the them fool a classifier under various different viewpoints.
In this exciting work, the 3D objects serve as a surface to carry information-rich and robust textures that can fool classifiers.
In our work, we also focus on perturbation on 3D objects, but we explicitly suppress the effect of textures by starting from 3D objects~\citep{xiang2014beyond} that have constant reflectance.
Even with constant reflectance, those 3D objects such as airplanes, bicycles, are easily recognizable due to their distinctive 3D shape features.
In this way, we highlight the importance of these shape features of objects in adversarial attacks.

Beyond perturbations in texture form, \citet{zeng2017adversarial} perturbed the physical parameters (normal, illumination and material) for untargeted attacks against 3D shape classification and
a VQA system. However, for the differentiable renderer, they assume that the camera parameters are known beforehand and then perturb 2D normal maps under the fixed projection. This may limit the manipulation space and may also produce implausible shapes. For the non-differentiable renderer in their work, they have to use derivative-free optimization for attacks. In comparison, our method can generate plausible shapes directly in mesh representation using gradient based optimization methods.

A concurrent work~\citep{liu2018adversarial} proposes to manipulate lighting and geometry to perform attacks. However, there are several differences compared to our work: 1) \emph{Magnitude of perturbation.} The perturbation in \cite{liu2018adversarial} such as lighting change is visible, while we achieve almost unnoticeable perturbation which is important in adversarial behaviors.
2) \emph{Targeted attack.} Based on the objective function in~\citep{liu2018adversarial} and experimental results, the adversarial targets seem close to each other, such as jaguar and elephant. In our work, we explicitly force the object from each class to be targeted-attacked into all other classes with almost 100\% attack success rate. 3) \emph{Renderers.} We perform attacks based on the state-of-the-art open-source differentiable renderer~\citep{nmrrepo}, which makes our attacks more accessible and reproducible, while in~\citep{liu2018adversarial} a customized renderer is applied and it is difficult to tell whether such vulnerabilities come from the customized renderer or the manipulated object. 4) \emph{Realistic attacks.} Manipulating lighting is less realistic in open environments. Compared with their attacks on lighting and shape, we propose to manipulate shape and texture of meshes which are easier to conduct in practice. 
5) \emph{Victim learning models.} We attack both classifiers and object detectors, which is widely used in safety-sensitive applications such as autonomous driving, while they only attack classifiers.

\noindent\textbf{Differentiable Renderers}
Besides adversarial attacks, differentiable renderers have been used in many other tasks as well, including inverse rendering~\citep{barron2015sirfs,genova2018morphable}, 3D morphable face reconstruction~\citep{genova2018morphable}, texture optimization~\citep{mordvintsev2018differentiable} and so on~\citep{li2018differentiable}.
In these tasks, gradient based optimization can be realized due to readily available differentiable renderers~\cite{matthew2014opendr,kato2018renderer,genova2018morphable,nguyen-phuoc2018rendernet,ramamoorthi2001envmap,li2018differentiable}. We also used a differentiable renderer called Neural Mesh Renderer~\citep{kato2018renderer}, which is fast and can be integrated into deep neural networks effortlessly.

\paragraph{Watermarking for Meshes}
While mesh watermarking is also achieved by manipulating the meshes in a subtle way, the goal is different from ours: it is to hide secret data in the geometry by satisfying strict properties of vertices and edges~\cite{Praun1999Robust,Cayre2003data}; our task is to perturb
the mesh as long as the rendered image can fool a learning model while keeping the mesh perceptual realistic. On
the other hand, the challenges in developing 3D mesh watermarking helps to emphasize the challenges for our attack given the difficulties of generating perturbation in 3D domains.

\section{Problem Definition and Challenges}
In 2D domain, let $g$ be a machine learning model trained to map a 2D image $I$ to its category label $y$.
For $g$, an adversarial attacker targets to generate an adversarial image $I^\adv$ such that $g(I^\adv) \neq y$ (untargeted attack) or $g(I^\adv) = y'$ (targeted attack), where $y$ is the groundtruth label and $y'$ is our specified malicious target label.

Unlike adversarial attacks in 2D space, here the image $I$ is a rendered result of a 3D object $S$: $I=R(S; P, L)$, computed by a physically based renderer $R$ with camera parameters $P$ and illumination parameters $L$. In other words, it is not allowed to directly operate the pixel values of $I$, and one has to manipulate the 3D object $S$ to generate $S^\adv$ such that the rendered image of it will fool $g$ to make incorrect predictions: $I^\adv = R(S^\adv; P, L)$.
Achieving the above goals is non-trivial due to the following challenges:
1) \textbf{Manipulation space}: When rendering 3D contents, shape, texture and illumination are entangled together to generate the pixel values in a 2D image, so image pixels are no longer independent with each other. This means the manipulation space can be largely reduced due to image parameterization.
2) \textbf{Constraints in 3D}: 3D constraints such as physically possible shape geometry and texture are not directly reflected on 2D~\citep{zeng2017adversarial}. Human perception of an object are in 3D or 2.5D~\citep{marr1982vision}, and perturbation of shape or texture on 3D objects may directly affect human perception of them. This means it can be challenging to generate unnoticeable perturbations on 3D meshes.
\section{Methodology}
Here we assume the renderer $R$ is known (\ie white box) and differentiable to the input 3D object $S$ in mesh representation.
To make a renderer $R$ differentiable, we have to make several assumptions regarding object material, lighting models, interreflection \etc.
Please refer to supplementary material for more details on differentiable rendering and mesh representation.
With a differentiable renderer, we can use gradient-based optimization algorithms to generate the mesh perturbations in an end-to-end manner, and we denote this method \stAdv.
\subsection{Optimization Objective}
We optimize the following objective loss function with respect to $S^\adv$, given model $g$ and target label $y'$:
\begin{equation}\label{eq:objective}\small
    \gL(S^\adv; g, y') = \gL_{\adv}(S^{\adv}; g, y') + \lambda \gL_{\perc}(S^{\adv})
\end{equation}
In this equation, $\gL_\adv$ is the adversarial loss to fool the model $g$ into predicting a specified target $y'$ (\ie $g(I^\adv) = y'$), given the rendered image $I^\adv=R(S^\adv;P,L)$ as input.
$\gL_\perc$ is the loss to keep
the ``adversarial mesh'' perceptually realistic.
$\lambda$ is a balancing hyper-parameter.
We further instantiate $\gL_\adv$ and $\gL_\perc$ in the next subsections, regarding different tasks (classification or object detection)
and perturbation types (shape or texture).
\subsubsection{Adversarial Loss}
\paragraph{Classification}
For a classification model $g$, the output is usually the probability distribution of object categories,
given an image of the object as the input. We use the cross entropy loss~\citep{de2005tutorial} as the adversarial loss for \stAdv:
\begin{equation}\small
    \gL_{\adv}(S^{\adv}; g, y') = y'\log(g(I^{\adv})) + (1- y') \log (1 - g( I^{\adv} )),
\end{equation}
where $I^\adv = R(S^\adv; P, L)$, and $y'$ is one-hot representation of the target label.

\paragraph{Object Detection}
For object detection, we choose a state-of-the-art model \yolo~\citep{redmon2018yolov3} as our victim model.
It divides the input image $I$ into $Z\times Z$ different grid cells.
For each grid cell, \yolo predicts the locations and label confidence values of $B$ bounding boxes. For each bounding box, it generates 5 values (4 for the coordinates and 1 for the objectness score) and a probability distribution over $N$ classes.
Here the adversary's goal is to make the victim object disappear from the object detector, called \emph{disappearance attack}.
So we use the disappearance attack loss~\citep{eykholt2018physical} as our adversarial loss for \yolo:
\begin{equation}\small
    \gL_{\adv}(S^\adv; g, y') = \max_{z\in Z^2, b\in B} H(z, b, y', g(I^{\adv})),
\end{equation}
where $I^\adv = R(S^\adv; P, L)$,
and $H(\cdot)$ is a function to represent the probabilities of label $y'$ for bounding box $b$ in the grid cell $z$, given $I^\adv$ as input of model $g$.

\subsubsection{Perceptual Loss}
To keep the ``adversarial mesh'' perceptually realistic, we leverage
a Laplacian loss similar to the total variation loss~\citep{vogel1996iterative} as our perceptual loss:
\begin{equation}\label{eq:adv-t}\small
\gL_{\perc}(S^\adv)=\sum_{i}\sum_{q\in\mathcal{N}(i)}\norm{I_i^\adv - I_q^\adv}_2^2,
\end{equation}
where $I_i$ is the RGB vector of the $i$-th pixel in the image $I^\adv=R(S^\adv; P, L)$, and $\mathcal{N}(i)$ is the 4-connected neighbors of pixel $i$.

We apply this smoothing loss to the image when generating texture based perturbation for $S^\adv$.
However, for shape based perturbation, manipulation of vertices may introduce unwanted mesh topology change,
as reported in~\citep{kato2018renderer}. Therefore, instead of using Eq.~\refeqshort{adv-t}, we perform smoothing on the displacement of vertices such that neighboring vertices
will have similar displacement flow. We achieve this by extending the smoothing loss to 3D vertex flow, in the form of a Laplacian loss:
\begin{equation}\label{eq:adv-v}\small
   \gL_{\perc}(S^\adv) =  \sum_{ \vv_i \in V} \sum_{\vv_q \in{\mathcal{N}(\vv_i)}} \norm{\Delta\vv_i - \Delta\vv_q}^2_2,
\end{equation}
where $\Delta\vv_i=\vv_i^\adv - \vv_i$
is the displacement of the perturbed vertex $\vv_i^\adv$ from its original position $\vv_i$ in the pristine mesh,
and $\mathcal{N}(\vv_i)$ denotes connected neighboring vertices of $\vv_i$ defined by mesh triangles.

\section{Transferability to Black-Box Renderers}\label{sec:transferability}
Our \stAdv aims to white-box-attack the system $g(R(S; P, L))$ by optimizing $S$ end-to-end since $R$ is differentiable.
However, we hope to examine the potential of \stAdv for 3D objects in practice, where the actural renderer may be unavailable.

We formulate this as a black-box attack against a non-differentiable renderer $R'$ under unknown rendering parameters $P^\unknown,L^\unknown$, \ie we have limited access to $R'$ but we still want to generate $S^\adv$ such that $R'(S^\adv, P^\unknown, L^\unknown)$ fools the model $g$.
Because we have no assumptions on the black-box renderer $R'$, it can render photorealistic images at a high computational cost, by enabling interreflection, occlusion and rich illumination models \etc such that the final image is an accurate estimate under real-world physics as if captured by a real camera.
In this case, the transferability of ``adversarial meshes'' generated by \stAdv is crucial since we want to avoid the expensive computation in $R'$ and still be able to generate such $S^\adv$.

We analyze two scenarios for such transferability.

\paragraph{Controlled Rendering Parameters}
Before black-box attacks, we want to first test our ``adversarial meshes'' directly under
the same rendering configuration (lighting parameters $L$, camera parameters $P$), only replacing the
the differentiable renderer $R$ with the photorealistic renderer $R'$.
In other words, while $I^\adv=R(S^\adv; P, L)$ can fool the model $g$ as expected,
we would like to see whether $I'^\adv=R'(S^\adv; P, L)$ can still fool the model $g$.
\paragraph{Unknown Rendering Parameters}
In this scenario, we would like to use \stAdv to attack a non-differentiable system $g(R'(S; P^\unknown, L^\unknown))$ under fixed, unknown rendering parameters $P^\unknown, L^\unknown$.
In practice, we will have access to the mesh $S$ and its mask $M$ in the original photorealistic rendering $I'=R'(S; P^\unknown, L^\unknown)$, as well as the model $g$.
Directly transfer from one renderer to another may not work due to complex rendering conditions.
To improve the performance of such black-box attack, we propose a pipeline as follows:
\begin{enumerate}
\item Estimate camera parameters $\hat{P}$ by reducing the error of object silhouette
        $\norm{R_\mathrm{mask}(S; P) - M}^2$, where $R_\mathrm{mask}(S; P)$ renders the mask of the object $S$ (lighting is irrelevant to produce the mask);
\item Given $\hat{P}$, estimate lighting parameters $\hat{L}$ by reducing the masked error of rendered images: $\norm{M\circ(R(S; \hat{P}, L)-I')}^2$, where the operator $\circ$ is Hadamard product;
\item Given $\hat{P}, \hat{L}$, use \stAdv to generate the ``adversarial mesh'' $S^\adv$ such that $R(S^\adv; \hat{P}, \hat{L})$ fools $g$; To improve robustness, we add random perturbations to $\hat{P}$ and $\hat{L}$ when optimizing;

\item Test $S^\adv$ in the original scene with photorealistic renderer $R'$: obtain the prediction $g(R'(S^\adv; P^\unknown, L^\unknown))$.
\end{enumerate}

\iftrue
\begin{table*}[thb]
    \centering
    \begin{small}
    \begin{adjustbox}{max width=\linewidth}
    \begin{tabular}{ccccccccc}
        \toprule
    \multirow{2}{*}{\shortstack{  Perturbation\\ Type}} & \multirow{2}{*}{Model}  & \multirow{2}{*}{\shortstack{Test \\Accuracy}} &\multicolumn{2}{c}{ Best Case} &\multicolumn{2}{c}{ Average Case} & \multicolumn{2}{c}{ Worst Case} \\ 
    \cmidrule(l{2pt}r{2pt}){4-5}\cmidrule(l{2pt}r{2pt}){6-7}\cmidrule(l{2pt}r{2pt}){8-9}
    & & & Avg.~Distance  & Succ.~Rate & Avg.~Distance & Succ.~Rate & Avg.~Distance & Succ.~Rate\\  \midrule
    \multirow{2}{*}{Shape}&DenseNet & $100.0\%$ & $8.4\times10^{-5}$ & $100.0\%$ & $1.8\times 10^{-4}$ & $100.0\%$ & $3.0\times 10^{-4}$ & $100.0\%$ \\ 
    &Inception-v3 & $100.0\%$ & $4.8\times 10^{-5}$ & $100.0\%$ & $1.2\times 10^{-4}$ & $\phantom{0}99.8\%$ & $2.3\times 10^{-4}$ & $\phantom{0}98.6\%$ \\
    \midrule
    \multirow{2}{*}{Texture}&DenseNet & $100.0\%$ & $3.8\times 10^{-3}$ & $100.0\%$ & $1.1\times 10^{-2}$ & $\phantom{0}99.8\%$ & $2.6\times 10^{-2}$ & $\phantom{0}98.6\%$ \\
    &\inception & $100.0\%$ & $3.7\times 10^{-3}$ & $100.0\%$ & $1.3\times 10^{-2}$ & $100.0\%$ & $3.2\times 10^{-2}$ & $100.0\%$ \\
        \bottomrule
    \end{tabular}
    \end{adjustbox}
    \end{small}
    \caption{Attack success rate of \stAdv and average distance of generated perturbation for different models and different perturbation types. We choose rendering configurations in \emph{\pascal renderings} such that the models have 100\% test accuracy on pristine meshes so as to confirm the adversarial effects. The average distance for shape based perturbation is computed using the 3D Laplacian loss from \refeqlong{adv-v}. The average distance for texture based perturbation is the root-mean-squared error of face color change.}
    \label{tbl:adv} 
\end{table*}

\else

\begin{table}[b]
    \centering
    \begin{adjustbox}{max width=\linewidth}
    \begin{tabular}{c|c|cc|cc}
         \toprule
         \multicolumn{2}{c|}{\multirow{2}{*}{\shortstack{Perturbation\\ Type}}} & \multicolumn{2}{c|}{\multirow{2}{*}{Shape}} & \multicolumn{2}{c}{\multirow{2}{*}{Texture}} \\
         \multicolumn{2}{c|}{} & & & \\
         \midrule
         \multicolumn{2}{c|}{Model} & \densenet & \inception & \densenet & \inception \\
         \midrule
         \multicolumn{2}{c|}{Test Accuracy} & $100.0\%$ & $100.0\%$ & $100.0\%$ & $100.0\%$ \\
\iffalse
         \midrule
          \multirow{2}{*}{Best} & distance & $8.4\times 10^{-5}$ & $4.8\times10^{-4}$ & $3.8\times10^{-3}$ & $3.7\times10^{-3}$ \\
                                & prob & $100.0\%$ & $100.0\%$ & $100.0\%$ & $100.0\%$\\
          \midrule
          \multirow{2}{*}{Average} & distance & $1.8\times10^{-4}$ & $1.2\times10^{-4}$ & $1.1\times10^{-2}$ & $1.3\times10^{-2}$\\
                                & prob & $100.0\%$ & $\phantom{0}99.8\%$ & $\phantom{0}99.8\%$ & $100.0\%$\\
          \midrule
          \multirow{2}{*}{Worst} & distance & $3.0\times10^{-4}$ & $2.3\times10^{-4}$ & $2.6\times10^{-2}$ & $3.2\times10^{-2}$ \\
                                & prob & $100.0\%$ & $\phantom{0}98.6\%$ & $\phantom{0}98.6\%$ & $100.0\%$\\
\else
         \midrule
         \multirow{3}{*}{\shortstack{Average\\ Distance}} & Best & $8.4\times 10^{-5}$ & $4.8\times10^{-5}$ & $3.8\times10^{-3}$ & $3.7\times10^{-3}$ \\
                                  & Avg & $1.8\times10^{-4}$ & $1.2\times10^{-4}$ & $1.1\times10^{-2}$ & $1.3\times10^{-2}$ \\
                                  & Worst & $3.0\times10^{-4}$ & $2.3\times10^{-4}$ & $2.6\times10^{-2}$ & $3.2\times10^{-2}$ \\
         \midrule
         \multirow{3}{*}{\shortstack{Success\\ Rate}} & Best & $100.0\%$ & $100.0\%$ & $100.0\%$ & $100.0\%$ \\
                                  & Avg & $100.0\%$ & $\phantom{0}99.8\%$ & $\phantom{0}99.8\%$ & $100.0\%$ \\
                                  & Worst & $100.0\%$ & $\phantom{0}98.6\%$ & $\phantom{0}98.6\%$ & $100.0\%$ \\
\fi
         \bottomrule
    \end{tabular}
    \end{adjustbox}
    \caption{Attack success rate of \stAdv and average distance of generated perturbation for different models and different perturbation types. We filter rendering configurations in \emph{\pascal renderings} such that the models have 100\% test accuracy on pristine meshes.}
    \label{tbl:adv} 
\end{table}
\fi
\section{Experimental Results}
In this section, we first show the attack effectiveness of ``adversarial meshes" generated by \stAdv against classifiers under different settings.
We then visualize the perturbation flow of vertices to better understand the vulnerable regions of those 3D objects.
User studies show that the proposed perturbation is subtle and will not mislead human recognition.
In addition, we show examples of applying \stAdv to object detectors in physically realistic scenes.
Finally, we evaluate the transferability of ``adversarial meshes'' generated by \stAdv and illustrate how to use such transferability to attack a black-box renderer.
\subsection{Experimental Setup}

For victim learning models $g$, we choose \densenet~\citep{huang2017densely} and 
\inception~\citep{szegedy2016rethinking} trained on ImageNet~\citep{deng2009imagenet} %
for classification, and \yolo trained on COCO~\citep{lin2014microsoft} for object detection.
For meshes ($S$), we preprocess CAD models in \pascal~\citep{xiang2014beyond} using uniform mesh resampling
with MeshLab~\citep{cignoni2008meshlab} to increase triangle density. %
Since these 3D objects have constant texture values, for texture perturbation
we also start from constant as pristine texture.

For the differentiable renderer ($R$), we
use the off-the-shelf PyTorch implementation~\citep{paszke2017automatic,nmrrepo} of the Neural Mesh Renderer (NMR)~\citep{kato2018renderer} to generate ``adversarial meshes''.
For rendering settings ($R(\cdot; P, L)$) when attacking classifiers, we randomly sample camera parameters $P$ and lighting parameters $L$, and filter out configurations such that the classification models have 100\% accuracy when rendering pristine meshes. These rendering configurations are then fixed for evaluation, and we call meshes rendered under these configurations \emph{\pascal renderings}. In total, we have 7 classes, and for each class we generate 72 different rendering configurations. More details are shown in the supplementary material.

For optimizing the objective, we use Adam~\citep{kingma2014adam} as our solver.
In addition, we select the hyperparameter $\lambda$ in \refeqlong{objective} using binary search, with 5 rounds of
search and 1000 iterations for each round. 

\subsection{\textbf{\StAdv} on Classification}\label{sec:pascal_classification}
In this section, we evaluate quantitative and qualitative performance of \stAdv against classifiers. %

For each sample in our \emph{\pascal renderings}, we try to targeted-attack it into the other 6 categories.
Next, for each perturbation type (shape and texture) and each model (\densenet and \inception), we split the results into three different cases similar to~\cite{carlini2017towards}:
{\em Best Case} means we attack samples within one class to other classes and report on the target class that is \emph{easiest} to attack. {\em Average Case} means we do the same but report the performance on \emph{all} of the target classes. Similarly, {\em Worst case} means that we report on the target class that is \emph{hardest} to attack.
The corresponding results are shown in \reftbl{adv}, including attack success rate of \stAdv, and the evaluation on generated shape and texture based perturbation respectively. For shape based perturbation, we use the Laplacian loss from \refeqlong{adv-v} as the distance metric. For texture based perturbation, we compute the root-mean-square distance of texture values for each face of the mesh: $\sqrt{\frac1m\sum_{i=1}^m (\vt_i^\adv-\vt_i)^2}$, where $\vt_i$ is the texture color of the $i$-th face among the mesh's total $m$ faces.
The results show that \stAdv can achieve almost 100\% attack success rate for either adversarial perturbation types.

\begin{figure}[bt]
 \begin{subfigure}{.49\linewidth}
 \centerline{\small{Target class}}\vspace{-5px}
 \scriptsize{
  \rotatebox{45}{aeroplane}\hspace{-4pt}\rotatebox{45}{bicycle}\hspace{-3pt}\rotatebox{45}{boat}\hspace{3pt}\rotatebox{45}{bottle}\hspace{3pt}\rotatebox{45}{chair}\hspace{3pt}\rotatebox{45}{diningtable}\hspace{-10pt}\rotatebox{45}{sofa}\linebreak
  }
 \includegraphics[width=\linewidth]{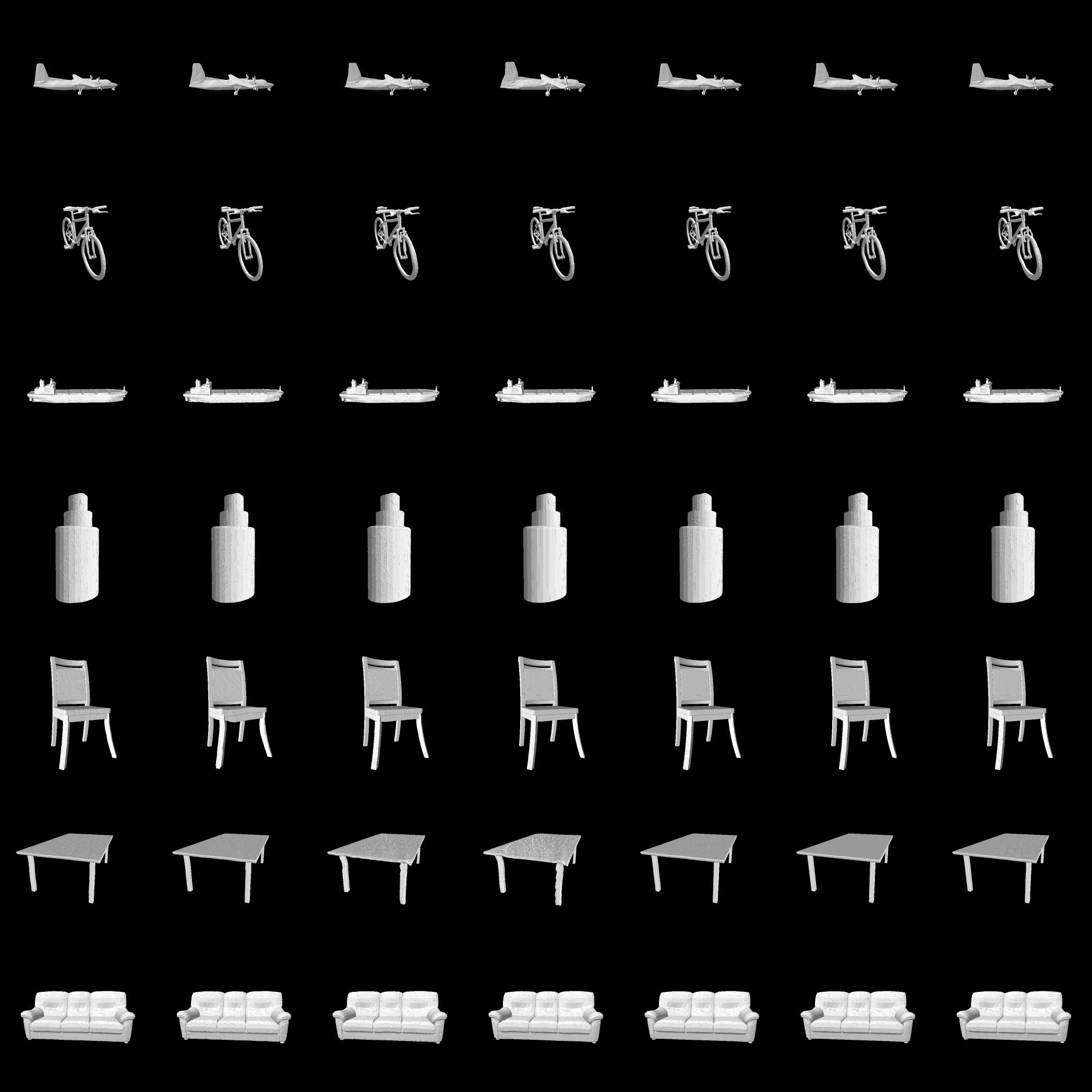}
 \caption{Perturbation on shape}
 \end{subfigure}
 \begin{subfigure}{.49\linewidth}
\centerline{\small{Target class}}\vspace{-5px}
 \scriptsize{
  \rotatebox{45}{aeroplane}\hspace{-4pt}\rotatebox{45}{bicycle}\hspace{-3pt}\rotatebox{45}{boat}\hspace{3pt}\rotatebox{45}{bottle}\hspace{3pt}\rotatebox{45}{chair}\hspace{3pt}\rotatebox{45}{diningtable}\hspace{-10pt}\rotatebox{45}{sofa}\linebreak
 }
 \includegraphics[width=\linewidth]{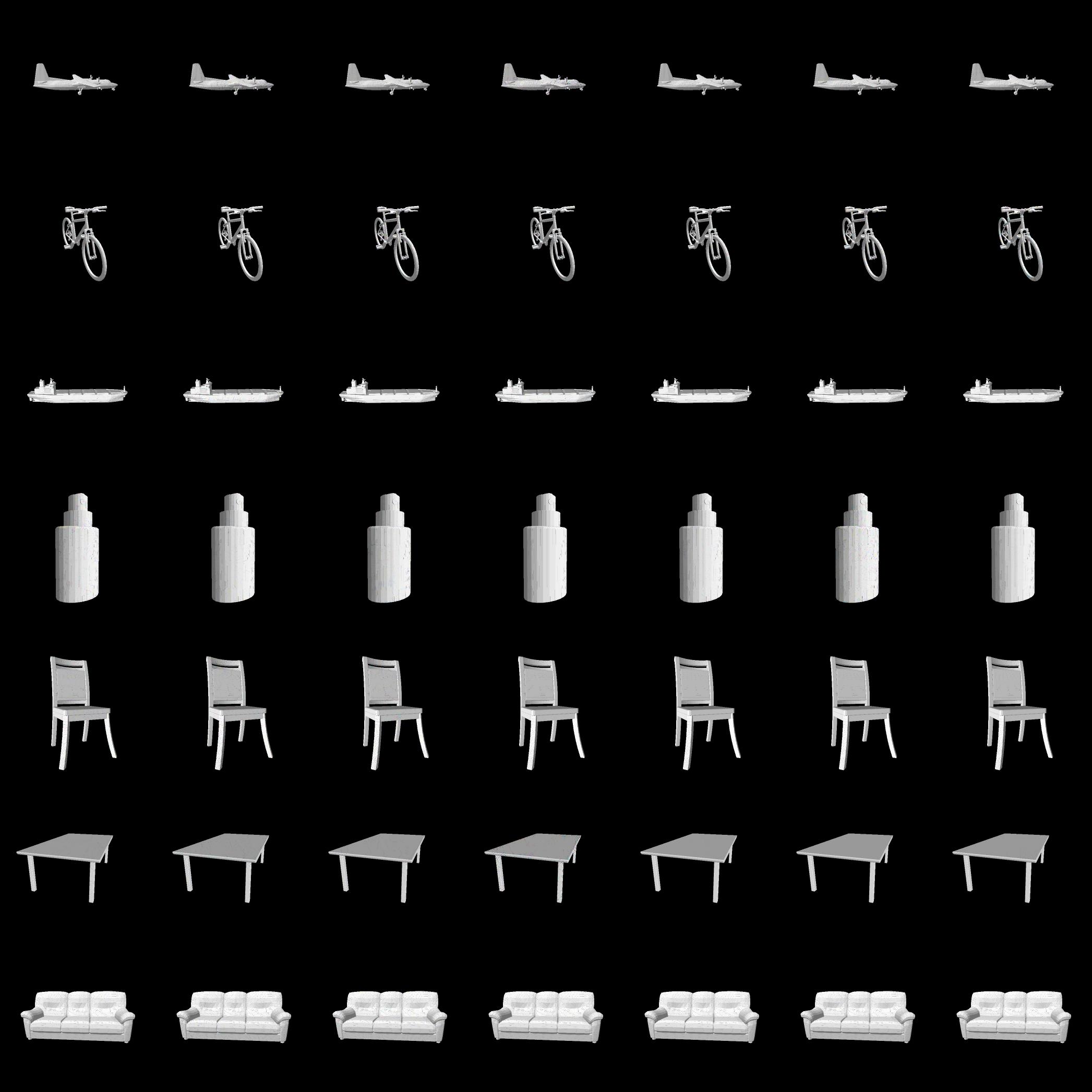}
 \caption{ Perturbation on texture}
 \end{subfigure}
\caption{Benign images (diagonal) and corresponding adversarial examples generated by \stAdv on \emph{\pascal renderings} tested on \inception. Adversarial target classes are shown at the top. We show perturbation on (a) shape and (b) texture. Similar results for \densenet can be found in the supplementary material.} %
\label{fig:adv-examples}
\end{figure}

\begin{figure}[t]
 \begin{subfigure}{.49\linewidth}
\centerline{\small{Target class}}\vspace{-5px}
  \scriptsize{
  \rotatebox{45}{aeroplane}\hspace{-4pt}\rotatebox{45}{bicycle}\hspace{-3pt}\rotatebox{45}{boat}\hspace{3pt}\rotatebox{45}{bottle}\hspace{3pt}\rotatebox{45}{chair}\hspace{3pt}\rotatebox{45}{diningtable}\hspace{-10pt}\rotatebox{45}{sofa}\linebreak
 }
 \includegraphics[width=\linewidth]{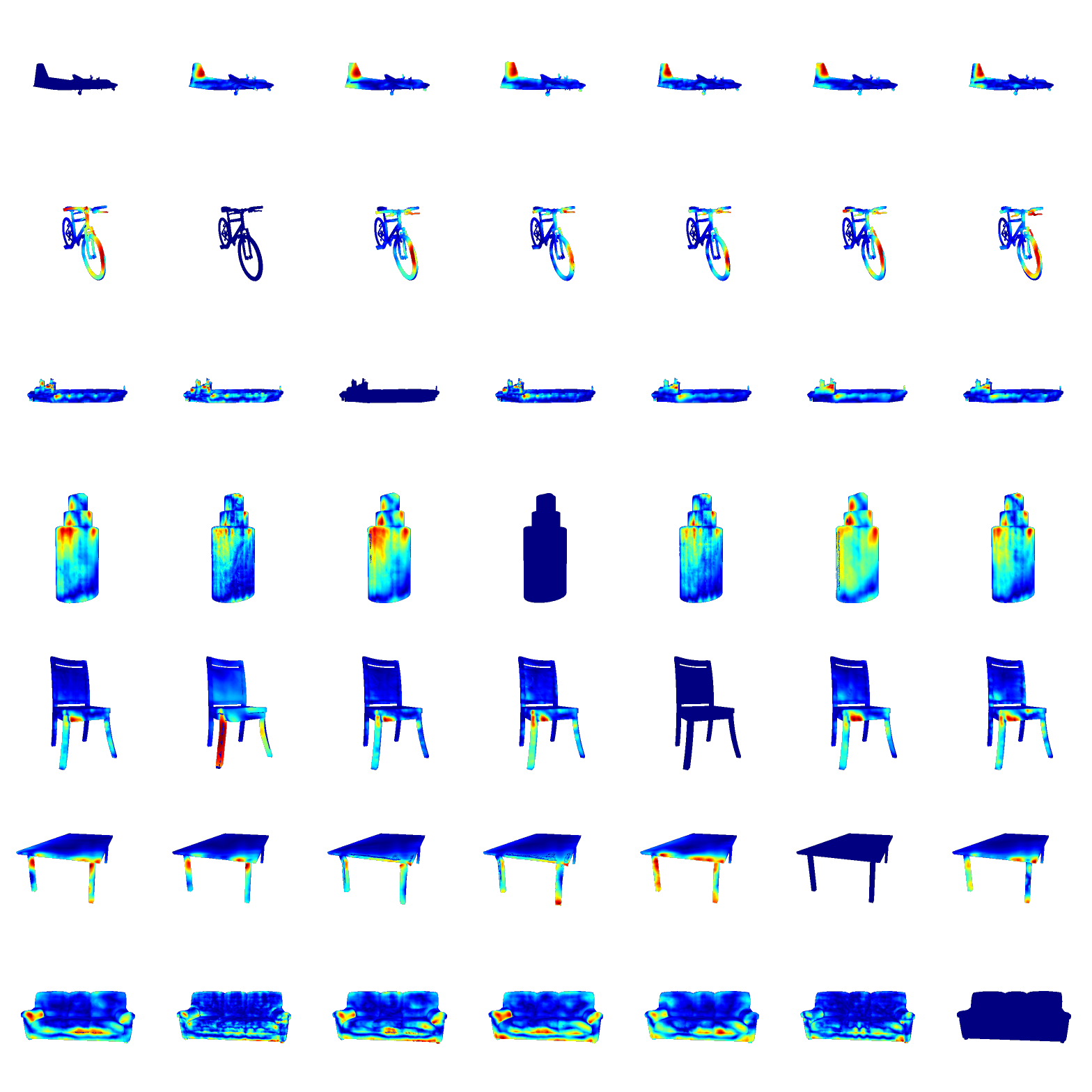}
  \caption{Rendered view}
 \end{subfigure}
 \unskip\hfill\vrule\hfill
\begin{subfigure}{.49\linewidth}
\centerline{\small{Target class}}\vspace{-5px}
 \scriptsize{
  \rotatebox{45}{aeroplane}\hspace{-4pt}\rotatebox{45}{bicycle}\hspace{-3pt}\rotatebox{45}{boat}\hspace{3pt}\rotatebox{45}{bottle}\hspace{3pt}\rotatebox{45}{chair}\hspace{3pt}\rotatebox{45}{diningtable}\hspace{-10pt}\rotatebox{45}{sofa}\linebreak
 }
    \includegraphics[width=\linewidth]{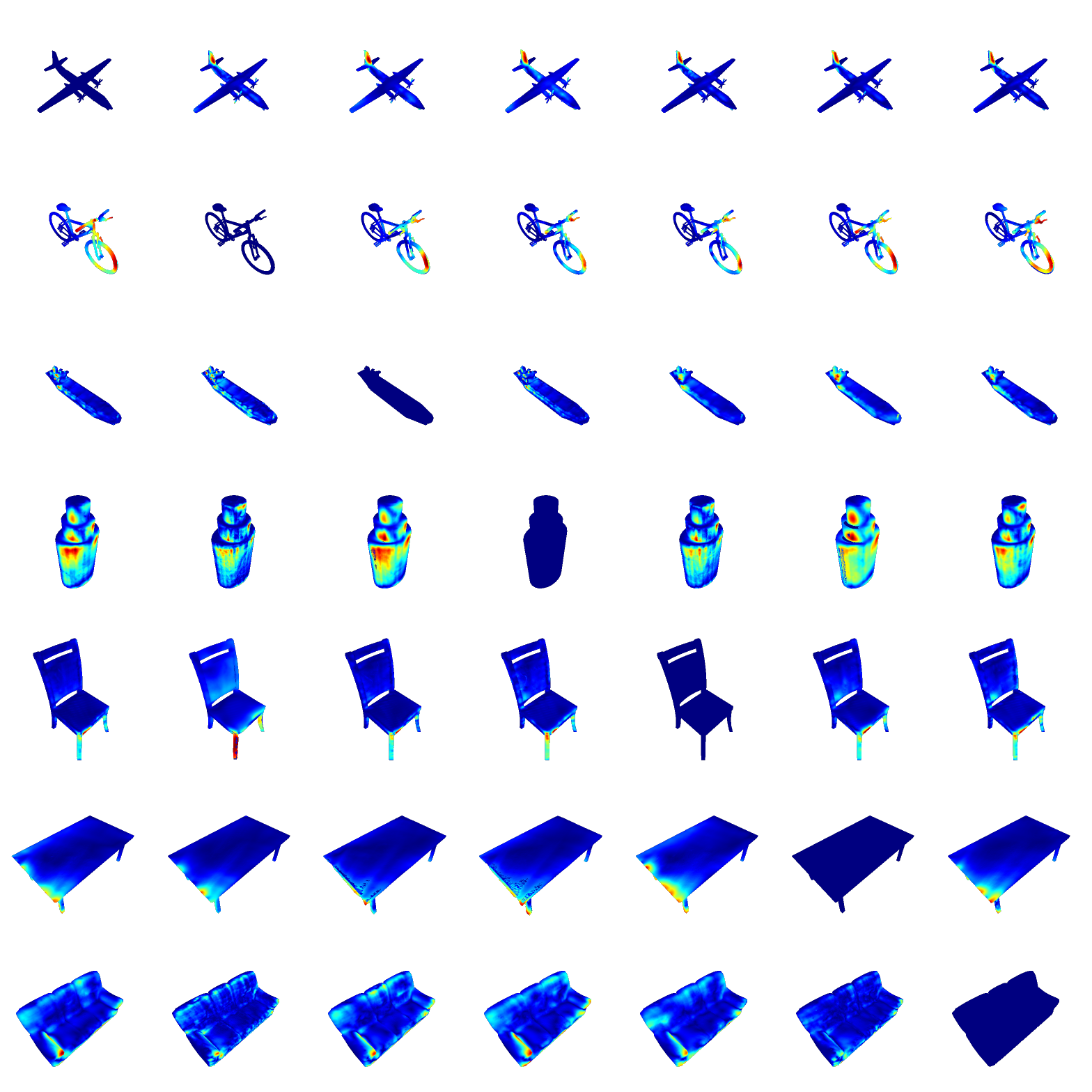}
      \caption{Canonical view}
\end{subfigure}
\vspace{5pt}

 \begin{subfigure}[t]{0.6\linewidth}
 \includegraphics[width=\linewidth]{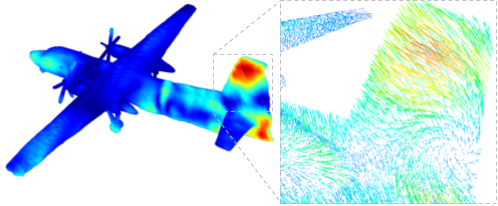}
      \caption{Vertex flow of an ``adversarial mesh'' targeting ``bicycle''}
    \label{fig:flow_zoomin}
 \end{subfigure}
 ~
 \begin{subfigure}[t]{0.35\linewidth}
 \includegraphics[width=\linewidth]{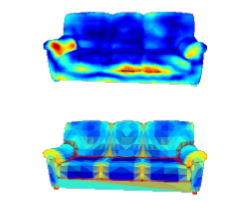}
 \caption{Perturbation (top) vs curvature (bottom)}
 \lblfig{mag_curv}
 \end{subfigure}

\caption{(a) and (b) are visualization of shape based perturbation with respect to \reffig{adv-examples}(a). (c) is a close view of flow directions, and (d) is an example to compare the magnitude of perturbation with the magnitude of curvature. Warmer color indicates greater magnitude and vice versa.}
\label{fig:flow-heatmap}
\end{figure}

\reffig{adv-examples} shows the generated ``adversarial meshes''
against \inception
after manipulating the vertices and texture respectively. The diagonal shows the images rendered with the pristine meshes.
The target class of each ``adversarial
mesh'' is shown at the top, and similar results for \densenet are included in the supplementary material.
Note that the samples in the image are randomly selected and not manually curated.
It is worth noting that the perturbation on object shape or texture, generated by \stAdv, is barely noticeable to human, while being able to mislead classifiers.
To help assess the vertex displacement in shape perturbation, we discuss the flow visualization and human perceptual study in the following paragraphs.
\paragraph{Visualizing Vertex Manipulation}
In order to better understand the vulnerable regions of 3D objects,
in \reffig{flow-heatmap}, we visualize the magnitude of the vertex
manipulation flow using heatmaps. The heatmaps in the figure correspond to the ones in \reffig{adv-examples}(a).
We adopt two viewpoints in this figure: the rendered view (i), which is the same as the one used for
rendering the images; and the canonical view (ii), which is achieved by fixing camera parameters for all shapes: we set the azimuth angle to $135^\circ$ and the elevation angle to $45^\circ$.
From the heatmaps we observe that the regions with large curvature value and close to the camera (such as edges)
are more vulnerable, as shown in the example in \reffig{flow-heatmap}(d).
We find this is reasonable, since vertex displacement in those regions will bring significant change
to normals, thus affecting the shading from the light sources and causing the screen pixel value
to change drastically.

In addition to magnitude, we additionally show an example of flow directions in
\reffig{flow-heatmap}(c), which is a close-up 3D quiver plot of the vertex flow in the vertical
stabilizer region of an aeroplane. In this example, the perturbed aeroplane mesh
is classified to ``bicycle'' in its rendering.
From this figure, we observe that the adjacent vertices tend to flow towards similar directions,
illustrating the effect of our 3D Laplacian loss operated on vertex flows (\refeqlong{adv-v}).

\paragraph{Human Perceptual Study}
We conduct a user study on Amazon Mechanical Turk (AMT) in order to quantify the realism of the adversarial meshes generated by \stAdv. We uploaded the adversarial images which are misclassified by \densenet and \inception. Participants were asked to recognize those adversarial object to one of the two classes (the ground-truth class and the adversarial target class).
The order of these two classes was randomized and the adversarial objects are appeared for 2 seconds in the middle of the screen during each trial. After disappearing, the participant has unlimited time to select the more feasible class according to their perception.
For each participant, one could only conduct at most 50 trials,
and each adversarial image was shown to 5 different participants.
The detailed settings of our human perceptual study are described in the supplementary material. In total, we collect 3820 annotations from 49 participants. In $99.29\pm 1.96 \%$ of trials the ``adversarial meshes'' were recognized correctly, indicating that our adversarial perturbation will not mislead human as they can almost always assign the correct label of these ``3D adversarial meshes''.

\paragraph{Multiview Robustness Analysis}
In addition to a fixed camera when applying \stAdv, we also explore the robustness of \stAdv against a range of viewpoints for shape based perturbation.
First, we create a victim set of images rendered under 5, 10 or 15 different azimuth angles for optimizing the attack. We then sample another 20 unseen views within the range for test. The results are shown in \reftbl{multiview}.
We can see that the larger the azimuth range is, the harder to achieve high attack success rate. In the meantime,
\stAdv can achieve relatively high attack success rate when more victim instances are applied for training. As a result, it shows that the attack robustness can potentially be improved under various viewpoints by optimizing on large victim set.

\begin{table}[t]
    \centering
    \begin{adjustbox}{max width=\linewidth}
    \begin{tabular}{cccc}
        \toprule
        \multirow{2}{*}{Victim Set Size} & \multicolumn{3}{c}{Azimuth Range} \\
        \cmidrule(l{2pt}){2-4}
        &\quad $45^\circ\sim60^\circ$ \quad&\quad$35^\circ\sim70^\circ$ \quad&\quad$15^\circ\sim75^\circ$ \\
        \midrule
        $\phantom{0}5$ views & $67\%$ & $45\%$ & $28\%$ \\
        $10$ views & $73\%$ & $58\%$ & $38\%$ \\
        $15$ views & $79\%$ & $74\%$ & $48\%$ \\
        \bottomrule
    \end{tabular}
    \end{adjustbox}
    \caption{Targeted attack success rate for unseen camera views. We attack using 5, 10, or 15 views, and test with 20 unseen views in the same range.
   } 
    \label{tbl:multiview} 
\end{table}

\subsection{\textbf{\StAdv} on Object Detection}
For object detection, we use \yolo~\cite{redmon2018yolov3} as our target model.

\paragraph{Indoor Scene}

\begin{figure}[tbh]
\centering
 \begin{subfigure}{0.35\linewidth}
 \includegraphics[width=\linewidth]{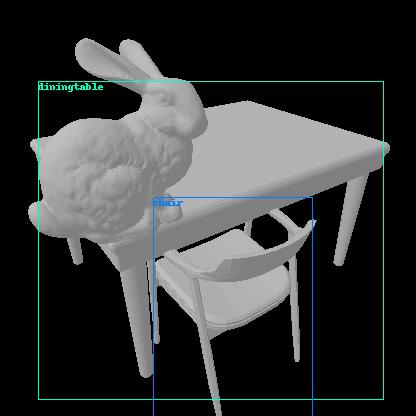}
 \caption{Benign}
 \end{subfigure}
 \unskip\hfill\vrule\hfill
 \begin{minipage}{0.60\linewidth}
 \begin{subfigure}{0.45\linewidth}
 \includegraphics[width=\linewidth]{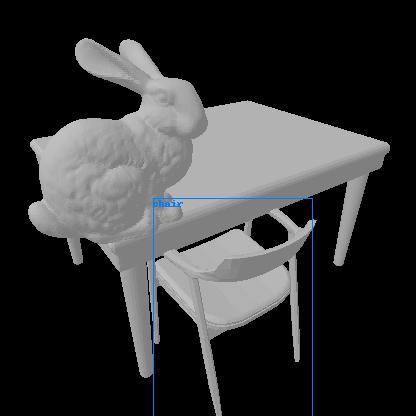}
 \caption{Table $\vert$ Shape }
 \end{subfigure}
 ~
 \begin{subfigure}{0.45\linewidth}
 \includegraphics[width=\linewidth]{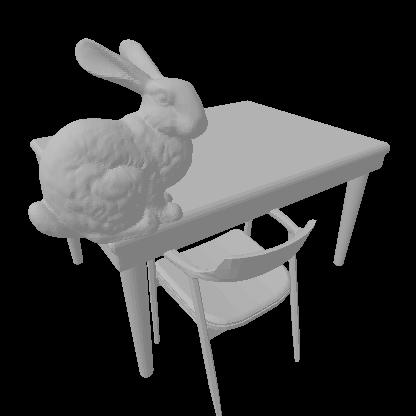}
 \caption{All $\vert$ Shape}
 \end{subfigure}

 \begin{subfigure}[t]{0.45\linewidth}
 \includegraphics[width=\linewidth]{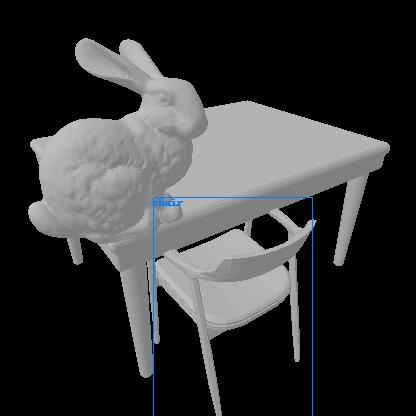}
 \caption{Table$\vert$Texture }
 \end{subfigure}
 ~
 \begin{subfigure}[t]{0.45\linewidth}
 \includegraphics[width=\linewidth]{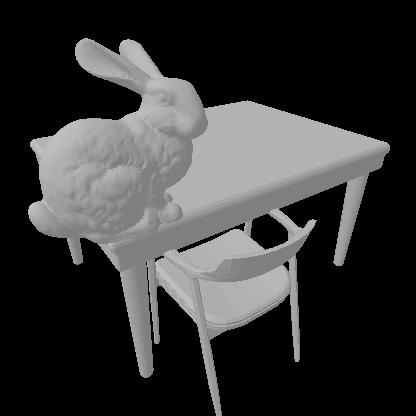}
 \caption{All $\vert$ Texture}
 \end{subfigure}
 \end{minipage}

\caption{``Adversarial meshes'' generated by \stAdv in a synthetic indoor scene. (a) represents the benign rendered image and (b)-(e) represent the rendered images from ``adversarial meshes'' by manipulating the shape or texture.
We use the format ``adversarial target $\vert$ perturbation type'' to denote the victim object aiming to hide and the type of perturbation respectively.  } \label{fig:attack-detector-indoor}
\vspace{-7px}
\end{figure}

First, we test \stAdv within the indoor scene which is pure synthetic. We compose the scene manually with a desk and a chair to simulate
an indoor setting, and place in the scene a single directional light with low ambient light.
We then put the Stanford Bunny mesh~\citep{turk1994bunny} onto the desk, and show that by
manipulating either the shape or the texture of the mesh, we can achieve the goal
of either removing the target table detection or removing all detections while keeping the perturbation almost  unnoticeable, as shown in~\reffig{attack-detector-indoor}.

\begin{figure}[t]
\centering
 \begin{subfigure}{.23\linewidth}
 \includegraphics[width=\linewidth]{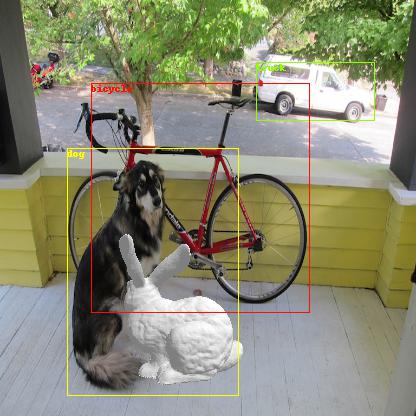}
 \caption{ $S$ $\vert$ GT }
 \end{subfigure}
 ~
 \begin{subfigure}{.23\linewidth}
 \includegraphics[width=\linewidth]{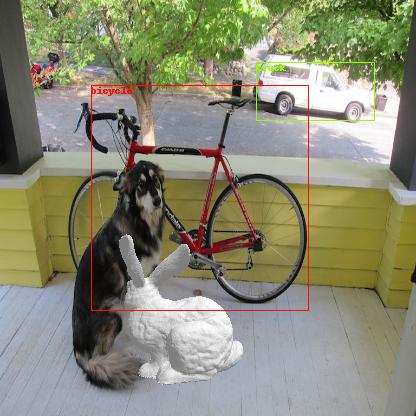}
 \caption{$S^\adv$ $\vert$ Dog}
 \end{subfigure}
 \begin{subfigure}{.23\linewidth}
 \includegraphics[width=\linewidth]{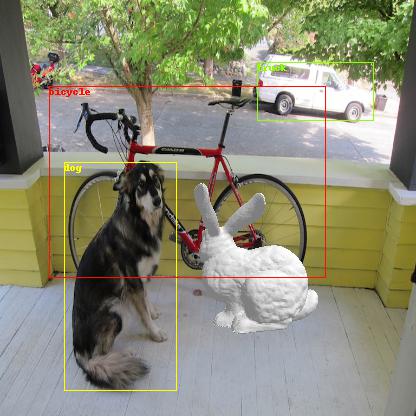}
 \caption{$S$ $\vert$ GT }
 \end{subfigure}
 ~
 \begin{subfigure}{.23\linewidth}
 \includegraphics[width=\linewidth]{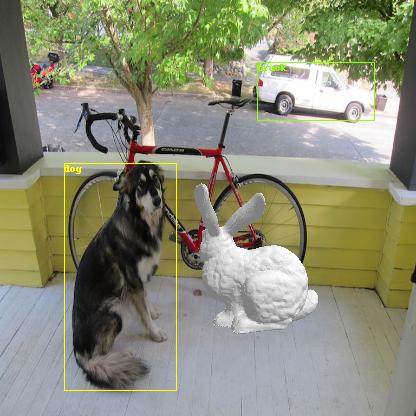}
 \caption{\scriptsize{$S^\adv\vert \textrm{Bicycle}$}}
 \end{subfigure}
\caption{``Adversarial meshes'' generated by \stAdv for an outdoor photo. (a) and (c) show images rendered with pristine meshes as control experiments, while (b) and (d) contain ``adversarial meshes'' by manipulating the shape. 
We use the format `` $S/S^\adv$ $\vert$ target''  to denote the benign/adversarial 3D meshes and the target to hide from the detector respectively.}
\label{fig:attack-detector-outdoor}
\vspace{-13pt}
\end{figure}

\paragraph{Outdoor Scene} %
Given a real photo of an outdoor scene, we hope to remove the detections of real objects in the photo.
Different from the indoor sceen in which lighting is known, we have to estimate the parameters of a sky lighting model~\citep{hosek2012sunsky}
using the API provided by \citet{holdgeoffroy2017outdoor} as groundtruth lighting and adapt to the differentiable renderer. We then use this lighting to render our mesh onto the photo.
In the real photo, we select the dog and the bicycle as our target objects and aim to remove the detection one at a time.
We show that we successfully achieve the adversarial goal with barely noticeable perturbation, as in \reffig{attack-detector-outdoor}.

\subsection{Transferability to Black-Box Renderers}
As mentioned in \refsec{transferability}, the final adversarial goal is to black-box attack a system $g(R'(S; P, L))$ in which the renderer $R'$ is a computationally intensive renderer that is able to produce photorealistic images. Here we choose Mitsuba~\citep{Mitsuba} as such renderer, and focus on shape based perturbation.

\begin{figure}
\centering
\includegraphics[width=\linewidth]{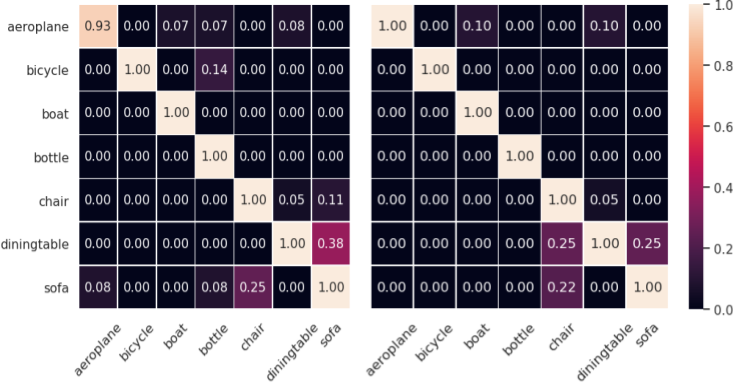}
\caption{Confusion matrices of targeted success rate for evaluating transferability of ``adversarial meshes'' on different classifiers. \textbf{Left}: \densenet; \textbf{right}: \inception.}\label{fig:confusion}
\vspace{-7px}
\end{figure}

\paragraph{Controlled Rendering Parameters}
Before perform such attacks, we first evaluate the transferability under controlled parameters.
We directly render the ``adversarial meshes'' $S^\adv$ generated in \refsec{pascal_classification} using Mitsuba, with the same lighting and camera parameters.
We then calculate the targeted/untargeted attack success rate by feeding the Mitsuba-rendered images to the same victim classification models $g$.
The result of untargeted attacks are shown in \reftbl{mitsuba-untarget}, and the confusion matrices for targeted attacks are show in \reffig{confusion}.
We observe that for untargeted attack, the ``adversarial meshes'' can be transferred to Mitsuba with relatively high atttack success rate for untargeted attack; while as shown in \reffig{confusion}, the targeted attack barely transfers in this straightforward setting.
\iffalse
\begin{table}[ht]
\centering
\adjustbox{max width=\linewidth}{
    \begin{tabular}{c|cc}
    \toprule
        Target\textbackslash Model & DenseNet & Inception-v3 \\
        \midrule
        aeroplane & $65.2\%$ & $67.1\%$ \\
        bicycle & $69.1\%$ & $83.8\%$ \\
        boat & $66.7\%$ & $39.6\%$ \\
        bottle & $63.0\%$ & $76.9\%$ \\
        chair & $37.1\%$ & $32.1\%$ \\
        diningtable & $70.3\%$ & $75.0\%$ \\
        sofa & $47.9\%$ & $52.3\%$ \\
        average & $59.8\%$ & $60.9\%$ \\
    \bottomrule
    \end{tabular}
}
    \caption{Transferability evaluation using untargeted attack success rate.}
    \label{tbl:mitsuba-untarget} 
\end{table}
We observe that untargeted attacks achieve high transferability, while that of target-attack is low except for some categories. 
\else
\begin{table}[t]
\setlength\aboverulesep{0pt}\setlength\belowrulesep{0pt}
\setcellgapes{3pt}\makegapedcells
    \centering
    \begin{adjustbox}{max width=\linewidth}
    \begin{tabular}{c|cccc}
        \toprule
     Model/Target & aeroplane & bicycle & boat  &bottle \\ 
     \midrule
    \densenet &  $65.2\%$ & $69.1\%$  & $66.7\%$ & $63.0\%$ \\ 
    \inception & $67.1\%$ & $83.3\%$ & $39.6\%$ & $76.9\%$ \\
    \midrule[\heavyrulewidth]
    Model/Target & chair & diningtable & sofa & \multicolumn{1}{|c}{\textbf{average}} \\
    \midrule
    \densenet    & $37.1\%$ & $70.3\%$ & $47.9\%$ & \multicolumn{1}{|c}{$59.8\%$}\\
    \inception   & $32.1\%$ & $75.0\%$ & $52.3\%$ & \multicolumn{1}{|c}{$60.9\%$}\\
    \bottomrule
    \end{tabular}
    \end{adjustbox}
    \caption{Untargeted attack success rate against Mitsuba by transferring ``adversarial meshes'' generated by attacking a differentiable renderer targeting different classes.
   } 
    \label{tbl:mitsuba-untarget} 
\end{table}
\fi

\paragraph{Unknown Rendering Parameters}

\begin{figure*}[htb]
    \centering
    \includegraphics[width=\linewidth]{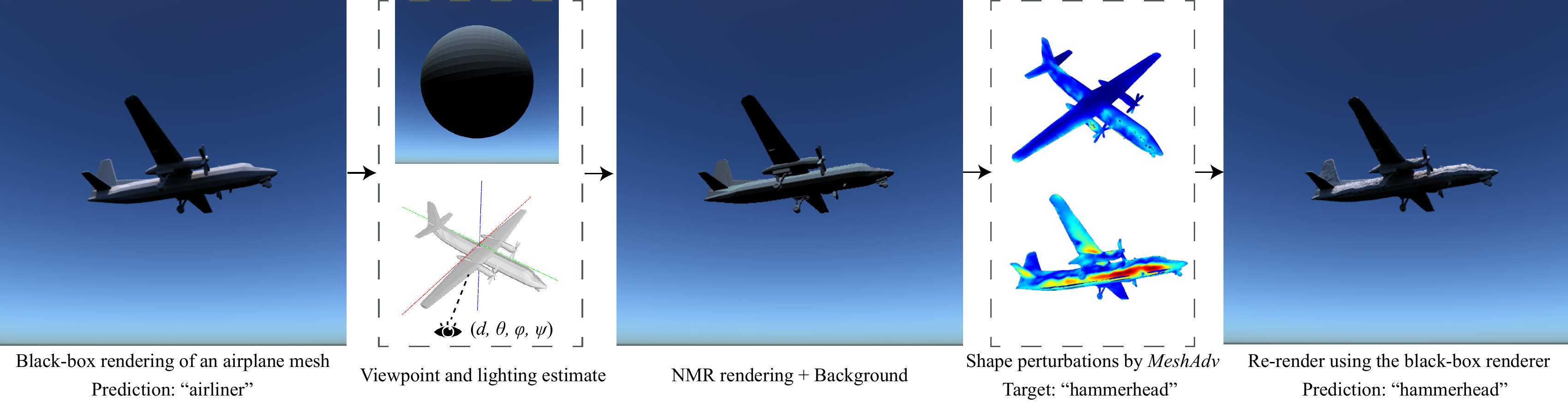}
    \caption{Transferability of ``adversarial meshes'' against classifiers in unknown rendering environment. We estimate the camera viewpoint and lighting parameters using the differentiable renderer NMR, and apply the generated ``adversarial mesh'' to the photorealistic renderer Mitsuba. The ``airliner'' is misclassified to the target class ``hammerhead'' after rendered by Mitsuba.}
    \label{fig:mitsuba_transfer_cls}
\end{figure*}

\begin{figure*}[tbh]
\label{bunny-obj}
\centering
 \begin{subfigure}{.19\linewidth}
 \includegraphics[width=\linewidth]{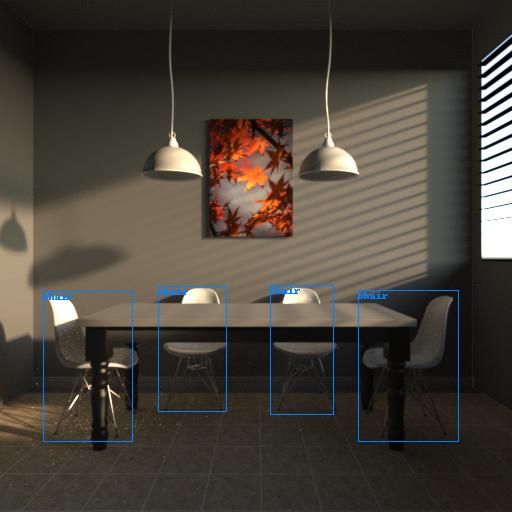}
 \caption{Benign}
 \end{subfigure}
 \unskip\hfill\vrule\hfill
 \begin{subfigure}{.19\linewidth}
  \begin{tikzpicture}[remember picture]
  \node[anchor=south west,inner sep=0] (imageA) {\includegraphics[width=\textwidth]{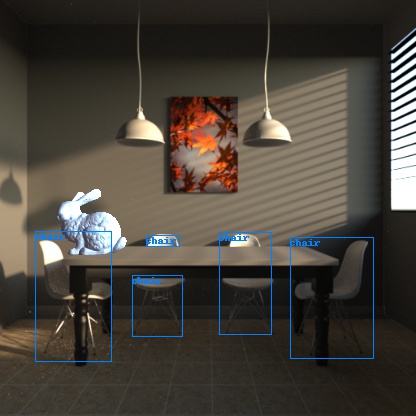}};
  \draw[->,ultra thick,red] (0.12,2.0) -- (0.32,1.5);
\end{tikzpicture}
 \caption{$S$ $\vert$ NMR }
 \end{subfigure}
 \begin{subfigure}{.19\linewidth}
  \begin{tikzpicture}[remember picture]
  \node[anchor=south west,inner sep=0] (imageA) {\includegraphics[width=\textwidth]{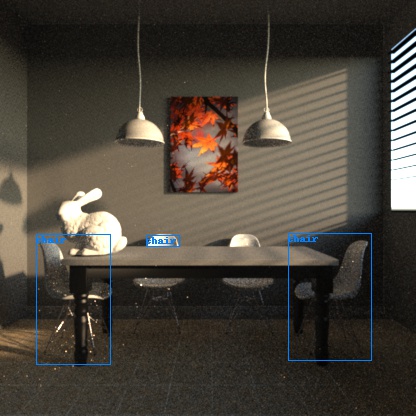}};
  \draw[->,ultra thick,red] (0.12,2.0) -- (0.32,1.5);
\end{tikzpicture}
 \caption{$S$ $\vert$ Mitsuba}
 \end{subfigure}
 \unskip\hfill\vrule\hfill
 \begin{subfigure}{.19\linewidth}
  \begin{tikzpicture}[remember picture]
  \node[anchor=south west,inner sep=0] (imageA) {\includegraphics[width=\textwidth]{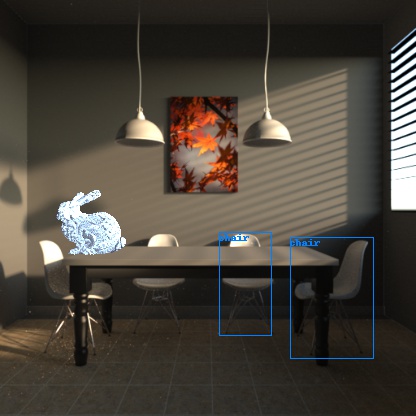}};
  \draw[->,ultra thick,red] (0.12,2.0) -- (0.32,1.5);
\end{tikzpicture}
 \caption{$S^\adv$ $\vert$ NMR }
 \end{subfigure}
 \begin{subfigure}{.19\linewidth}
  \begin{tikzpicture}[remember picture]
  \node[anchor=south west,inner sep=0] (imageA) {\includegraphics[width=\textwidth]{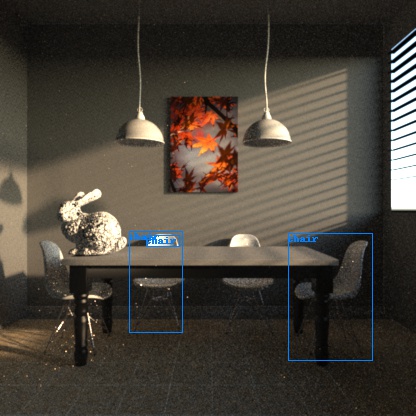}};
  \draw[->,ultra thick,red] (0.12,2.0) -- (0.32,1.5);
\end{tikzpicture}
 \caption{$S^\adv$ $\vert$ Mitsuba}
 \end{subfigure}
\caption{Transferability of ``adversarial meshes'' against object detectors in unknown rendering environment. (b) (c) are controlled experiments. $S^\adv$ is generated using NMR (d), targeting to hide the leftmost chair (see red arrows), and the adversarial mesh is tested on Mitsuba (e). We use ``$S/S^\adv$ $\vert$ renderer" to denote whether the added object is adversarially optimized and the renderer that we aim to attack with transferability respectively. }\label{fig:bunny-obj}
\vspace{-7px}
\end{figure*}
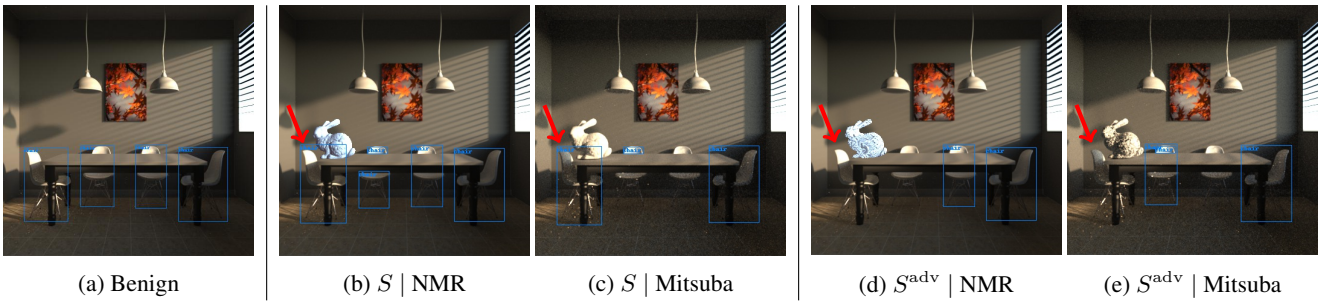

To more effectively targeted attack the system $g(R'(S; P^\unknown, L^\unknown))$ when rendering parameters $P^\unknown, L^\unknown$ are unknown,
we apply the pipeline from \refsec{transferability} on a classifier and an object detector, respectively.
we first use the Adam optimizer~\citep{kingma2014adam} to obtain the camera estimate $\hat{P}$, then 
estimate the lighting $L^\unknown$ using 5 directional lights and an ambient light $\hat{L}$.
Note that the groundtruth lighting $L^\unknown$ spatially varies due to interreflection and occlusion, so it is impossible to have an exact estimate using the global lighting model in NMR.
Then we manipulate the shape $S^\adv$ in the NMR until the image $I^\adv=R(S^\adv: \hat{P}, \hat{L})$
can successfully targeted-attack the classifier or the object detector $g$ with a high 
confidence. During this process, we add small random perturbation to the estimated parameters ($\hat{P}, \hat{L}$) such that $S^\adv$ will be more robust under uncertainties.
For testing, we re-render $S^\adv$ with Mitsuba using the original setting and test the rendered image $I'^\adv=R'(S^\adv, P^\unknown, L^\unknown)$ on the same model $g$.
For classification, we place an aeroplane object from \pascal and put it in an outdoor scene under sky light.
As is shown in \reffig{mitsuba_transfer_cls}, we successfully attacked the
classifier to output the target ``hammerhead'' by replacing the pristine mesh with
our ``adversarial mesh'' in the original scene. Note that even we do not have an accurate lighting estimate,
we still achieve the transferability by adding perturbation to lighting parameters.
For object detection, we modified a scene from \cite{bitterli2016rendering},
and placed the Stanford Bunny object into the scene.
The adversarial goal here is to remove the \emph{leftmost} chair in the image. Without an accurate lighting estimate, \reffig{bunny-obj} shows that the ``adversarial meshes'' can still successfully remove the target (the leftmost chair) from the detector.

\section{Conclusion}
In this paper, we proposed \stAdv to generate ``adversarial meshes'' by manipulating the shape or the texture of a mesh.
These ``adversarial meshes'' can be rendered to 2D domains to mislead different machine learning models. 
We evaluate \stAdv quantitatively and qualitatively using CAD models from \pascal, and also show that the adversarial behaviors of our ``adversarial meshes'' can transfer to black-box renderers.
This provides us a better understanding of adversarial behaviors of 3D meshes in practice, and can motivate potential future defenses.
\paragraph{Acknowledgement} 
We thank Lei Yang, Pin-Yu Chen for their valuable discussions on this work.
This work is partially supported by the
National Science Foundation under Grant CNS-1422211, CNS-1616575, IIS-1617767 and DARPA under Grant 00009970.
\newpage
{\small
\setlength{\bibsep}{0pt}
\bibliographystyle{abbrvnat}
\bibliography{main}
}

\renewcommand{\thesection}{\Alph{section}}
\renewcommand{\theequation}{S\arabic{equation}}
\renewcommand{\thefigure}{\Alph{figure}}
\onecolumn
\pagebreak
% \widetext
\begin{center}
\textbf{\Large Supplemental Material}
\end{center}
\setcounter{section}{0}
\setcounter{equation}{0}
\setcounter{figure}{0}
\setcounter{table}{0}
\setcounter{page}{1}

\section{Formulation of Differentiable Rendering }
A physically based renderer $R$ computes a 2D image $I=R(S; P, L)$
with camera parameters $P$, a 3D object $S$ and lighting parameters $L$ by
approximating physics, \eg the rendering equation~\citep{kajiya1986rendering,Immel1986radiosity}.
A differentiable renderer makes such
computation differentiable w.r.t.~the input $S, P, L$ by making assumptions on illumination models
and surface reflectance, and simplifying the ray-casting process. 
Following common practice, we use 3D  triangular  meshes for object representation, Lambertian surface for surface modeling, directional lighting with a uniform ambient
for illumination, and ignore interreflection and shadows.
Here, we further explain the details regarding 3D mesh representation $S=(V, F, T)$, illumination model $L$ and camera
parameters $P$ used in differentiable rendering in this work.

For a 3D object $S$ in 3D triangular mesh representation,
let $V$ be the set of its $n$ vertices in 3D space,
and $F$ be the indices of its $m$ faces:
\begin{equation}
    V = \{\vv_1, \vv_2, \cdots, \vv_n \in \sR^3\},\qquad F = \{\vf_1, \vf_2, \cdots, \vf_m \in \sN^3\}
\end{equation}
For textures, traditionally, they are represented by 2D texture images and mesh surface
parameterization such that the texture images can be mapped onto the mesh's triangles.
For simplicity, here we attach to
each triangular face a single RGB color as its reflectance:
\begin{equation}
T=\{\vt_1, \vt_2, \cdots, \vt_m \in {\sR^{+}}^3\}
\end{equation}

For illumination model, we use $k$ directional light sources plus an ambient light.
The lighting directions are denoted
$L_{\rm{dir}}$, and the lighting colors (in RGB color space) are denoted as $L_{\rm{color}}$ for directional
light sources and $\va$ for the ambient light:
\begin{equation}
L_{\rm{dir}} = \{\vl_1^d, \vl_2^d, \cdots \vl_k^d \in \sR^3\},\qquad L_{\rm{color}} =\{\vl_1^c, \vl_2^c, \cdots \vl_k^c \in \sR^3\}
\end{equation}

We put the mesh $S=(V, F, T)$ at the origin $(0, 0, 0)$, and set up our perspective camera 
following a common practice: the camera viewpoint is described by a quadruple $P=(d, \theta, \phi, \psi)$,
where $d$ is the distance of the camera to the origin, and $\theta$, $\phi$, $\psi$ are
azimuth, elevation and tilt angles respectively. Note that here we assume the camera intrinsics 
are fixed and we only need gradients for the extrinsic parameters $P$.

Given the above description, the 
2D image produced by the differentiable renderer can be symbolized as follows:

\begin{equation}
    I = \mathrm{rasterize}(P, T \cdot \mathrm{shading}(L, \mathrm{normal}(V, F)))
\end{equation}

\textit{normal($\cdot$, $\cdot$)} computes the normal direction $\vn_i$ for each triangular face $\vf_i$ in the mesh,
by computing the cross product of the vectors along two edges of the face:

\begin{equation}
    \vn_i = \frac{(\vv_{\vf_i^{(1)}} - \vv_{\vf_i^{(2)}}) \times (\vv_{\vf_i^{(2)}} - \vv_{\vf_i^{(3)}})}
    {\norm{(\vv_{\vf_i^{(1)}} - \vv_{\vf_i^{(2)}}) \times (\vv_{\vf_i^{(2)}} - \vv_{\vf_i^{(3)}})}_2}
\end{equation}

\textit{shading} computes the shading intensity $\vs_i$ on the face given the face normal direction $\vn_i$ and lighting parameters:

\begin{equation}
    \vs_i = \va + \sum_{i=1}^k \vl_i^c \max(\vl_i^d \cdot \vn_i, 0)
\end{equation}

Given face reflectance $\vt_i$ for each face $i$, we compute the color $\vc_i$ of each face $i$ by elementwise multiplication:

\begin{equation}
    \vc_i = \vt_i \circ \vs_i
\end{equation}

\textit{rasterize} projects the computed face colors $\vc_i$ in 3D space onto the 2D camera plane by raycasting and depth testing. We also cap the color values to $[0, 1]$.

For implementation, we use the off-the-shelf PyTorch implementation~\citep{paszke2017automatic,nmrrepo} of the Neural Mesh Renderer (NMR)~\citep{kato2018renderer}.

\section{\StAdv on Classification}
\paragraph{Creation of \emph{\pascal Renderings} }
For classification, we create \emph{\pascal renderings} using CAD models from \pascal~\cite{xiang2014beyond}. Those meshes are then scaled to $[-1,1]$ and put into the scene.
Then, we use Neural Mesh Renderer (NMR) to generate synthetic renderings using these unitized meshes with uniformly sampled random camera parameters: azimuth from $[0^\circ, 360^\circ)$, elevation from $[0^\circ, 90^\circ]$. As for lighting, we used a directional light and an ambient light for \emph{\pascal Renderings}. The direction is uniformly sampled in a cone such that the angle between the view and the lighting direction is less than $60^\circ$.

In order to obtain the groundtruth labels, we map the object classes in \pascal to the corresponding classes in the ImageNet. Next, we feed the synthetic renderings to \densenet and \inception and filter out the samples that are misclassified by
either network, so that both models have $100\%$ prediction accuracy on our \emph{\pascal renderings}. We then save the rendering configurations for evaluation of \stAdv.

\begin{figure*}[tbh]
\centering
\begin{minipage}{.45\textwidth}
\centerline{Target class}
 \begin{subfigure}{\textwidth}
  \rotatebox{45}{aeroplane}\hspace{0pt}\rotatebox{45}{bicycle}\hspace{10pt}\rotatebox{45}{boat}\hspace{10pt}\rotatebox{45}{bottle}\hspace{10pt}\rotatebox{45}{chair}\hspace{10pt}\rotatebox{45}{diningtable}\hspace{-5pt}\rotatebox{45}{sofa}\linebreak
 \includegraphics[width=\textwidth]{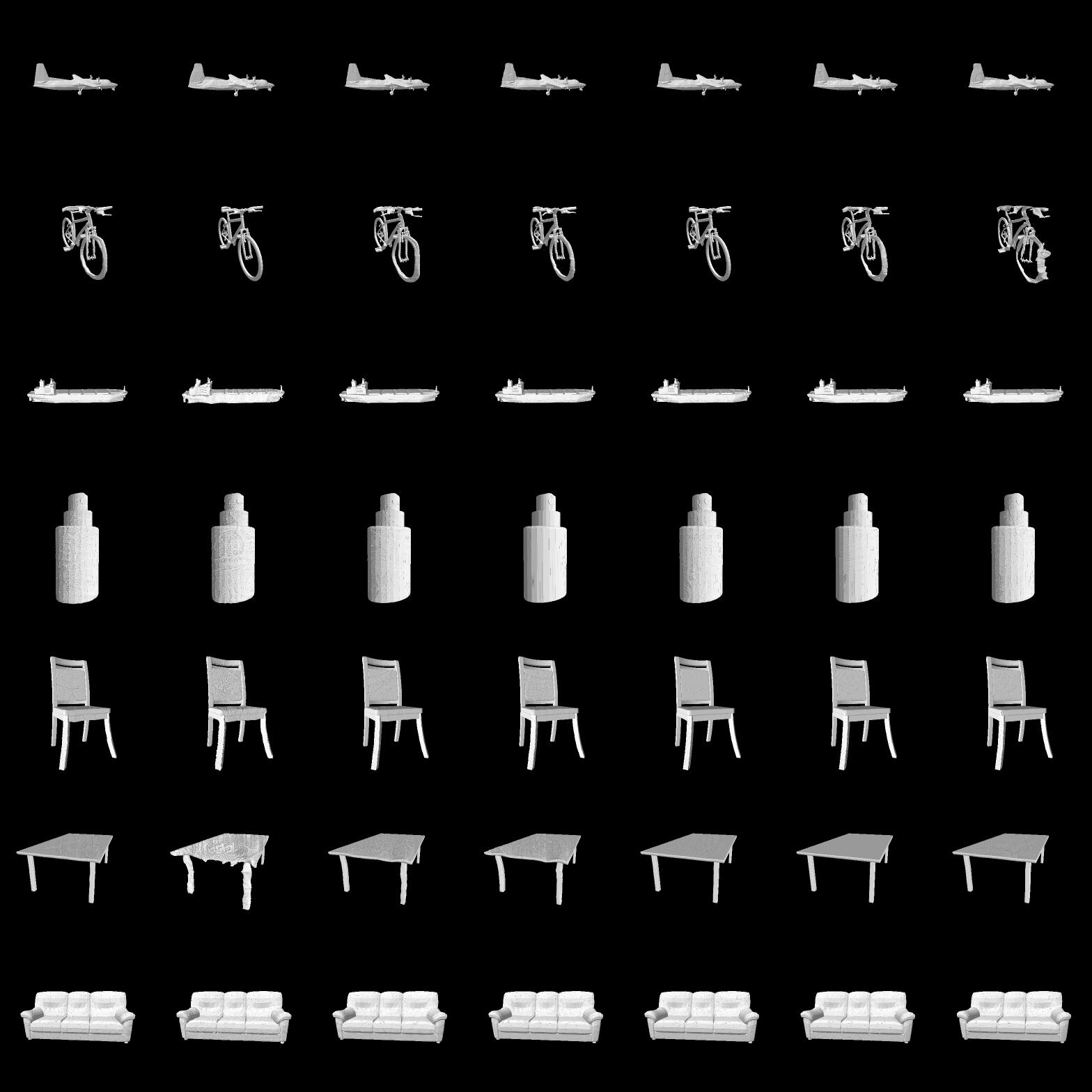}
 \caption{Perturbation on Shape}
 \end{subfigure}
\end{minipage}
\begin{minipage}{.45\textwidth}
\centerline{Target class}
 \begin{subfigure}{\textwidth}
  \rotatebox{45}{aeroplane}\hspace{0pt}\rotatebox{45}{bicycle}\hspace{10pt}\rotatebox{45}{boat}\hspace{10pt}\rotatebox{45}{bottle}\hspace{10pt}\rotatebox{45}{chair}\hspace{10pt}\rotatebox{45}{diningtable}\hspace{-5pt}\rotatebox{45}{sofa}\linebreak
 \includegraphics[width=\textwidth]{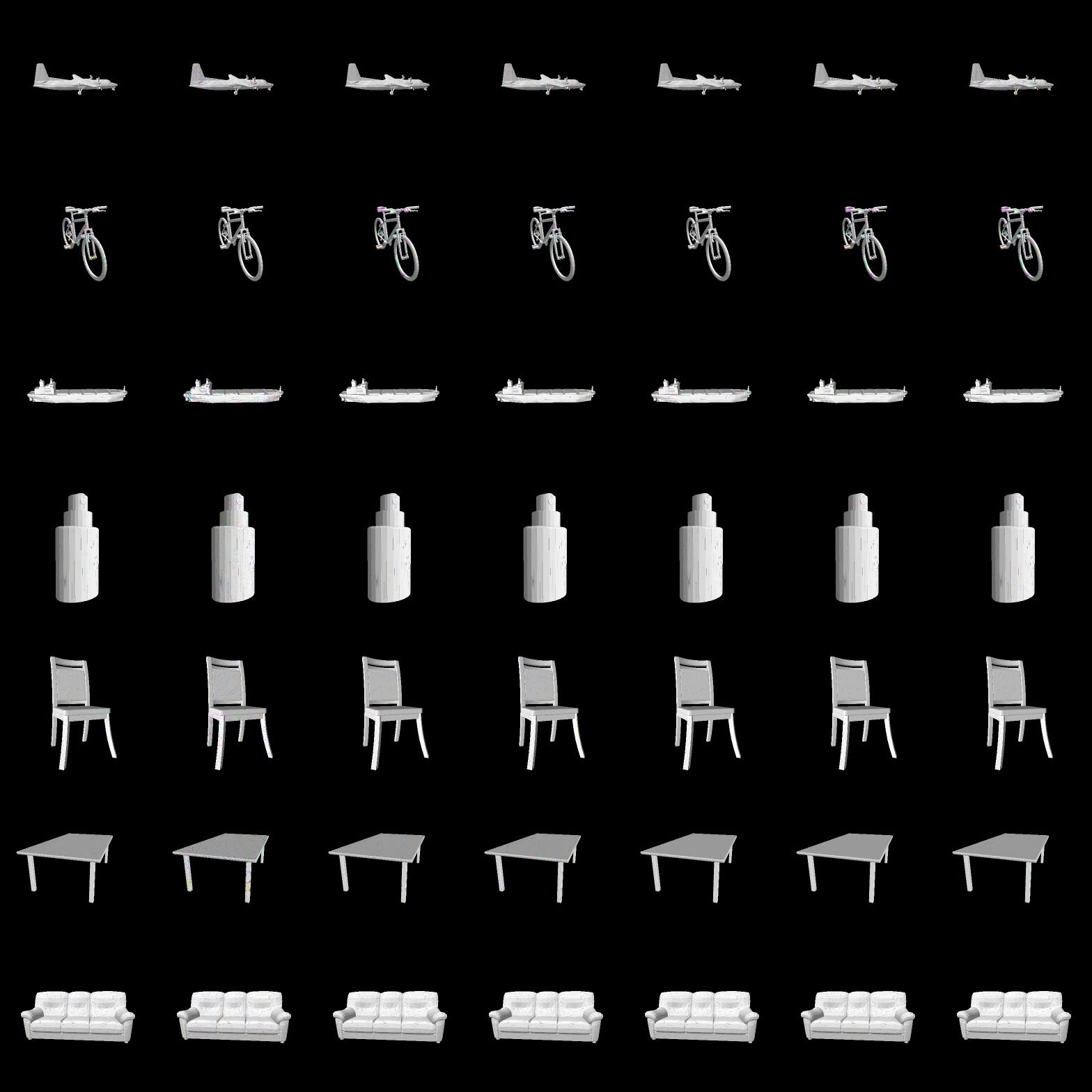}
 \caption{Perturbation on Texture}
 \end{subfigure}
\end{minipage}
\caption{Benign images (diagonal) and corresponding adversarial meshes generated by \stAdv on \pascal shapes against \densenet, targeting at different classes as shown on the top . (a) Presents the ``adversarial meshes'' by manipulating the shape; (b) by manipulating texture. }
\label{fig:inception-v3-result}
\end{figure*}

\paragraph{Additional Results for \densenet}
\reffig{inception-v3-result} shows the generated ``adversarial meshes'' against DenseNet, similar to Figure.~2 in the main paper.

\section{Human Perceptual Study Procedures}
We conduct a user study on Amazon Mechanical Turk (AMT) in order to quantify the realism of the ``adversarial meshes'' generated by \stAdv. We uploaded the adversarial images on which \densenet and \inception misclassify the object. Participants were asked to classify those adversarial images to one of the two classes (the groundtruth class and the target class).
The order of these two classes was randomized and the adversarial images appeared for 2 seconds in the middle of the screen on each trial. After disappearing, the participant had unlimited time to select the more feasible class according to her perception.
For each participant, one could only conduct at most 50 trials,
and each adversarial image was shown to 5 different participants.

\end{document}